\documentclass[aps,prb,twocolumn,showpacs,10pt,superscriptaddress,floatfix,longbibliography]{revtex4-1}

\usepackage{mathptmx}
\usepackage{graphicx}
\usepackage{subfigure}
\usepackage{calrsfs}
\DeclareMathAlphabet{\pazocal}{OMS}{zplm}{m}{n}
\usepackage{amsmath}
\usepackage{amsfonts}
\usepackage{amssymb}
\usepackage{mathrsfs}
\usepackage{bm}
\usepackage{subfigure}
\usepackage{xcolor}
\usepackage[colorlinks=true,linkcolor=blue,anchorcolor=red,citecolor=blue, urlcolor=blue]{hyperref}\usepackage{ulem}
\usepackage[bottom]{footmisc}

\usepackage{braket}

\newlength{\seplinewidth}
\newlength{\seplinesep}
\colorlet{sepline}{orange}

\begin{document}

\title{Prediction of Spin Polarized Fermi Arcs in Quasiparticle Interference of CeBi}

\author{Zhao Huang}
\affiliation{Theoretical Division, Los Alamos National Laboratory, Los Alamos, New Mexico 87545, USA}

\author{Christopher Lane}
\affiliation{Theoretical Division, Los Alamos National Laboratory, Los Alamos, New Mexico 87545, USA}
\affiliation{Center for Integrated Nanotechnology, Los Alamos National Laboratory, Los Alamos, New Mexico 87545, USA}

\author {Chao Cao}
\affiliation{Department of Physics, Hangzhou Normal University, Hangzhou 310036, China}

\author {Guo-Xiang Zhi}
\affiliation{Department of Physics, Zhejiang University, Hangzhou 310013, China}

\author{Yu Liu}
\affiliation{Department of Physics, Harvard University, Cambridge, Massachusetts 02138, USA}

\author{Christian E. Matt}
\affiliation{Department of Physics, Harvard University, Cambridge, Massachusetts  02138, USA}

\author{Brinda Kuthanazhi}
\affiliation{Ames Laboratory, Iowa State University, Ames, Iowa 50011, USA}
\affiliation{Department of Physics and Astronomy, Iowa State University, Ames, Iowa 50011, USA}

\author{Paul C. Canfield}
\affiliation{Ames Laboratory, Iowa State University, Ames, Iowa 50011, USA}
\affiliation{Department of Physics and Astronomy, Iowa State University, Ames, Iowa 50011, USA}

\author{Dmitry Yarotski}
\affiliation{Center for Integrated Nanotechnology, Los Alamos National Laboratory, Los Alamos, New Mexico 87545, USA}

\author{A. J. Taylor}
\affiliation{Center for Integrated Nanotechnology, Los Alamos National Laboratory, Los Alamos, New Mexico 87545, USA}

\author{Jian-Xin Zhu}
\email{jxzhu@lanl.gov}
\affiliation{Theoretical Division, Los Alamos National Laboratory, Los Alamos, New Mexico 87545, USA}
\affiliation{Center for Integrated Nanotechnology, Los Alamos National Laboratory, Los Alamos, New Mexico 87545, USA}

\begin{abstract}
We predict that CeBi in the ferromagnetic state is a Weyl semimetal. Our calculations within density functional theory show  the existence of  two pairs of Weyl nodes on the momentum path $(0, 0, k_z)$ at $15\; {\rm meV}$ above and $100\; {\rm meV}$ below the Fermi level. Two corresponding Fermi arcs are obtained on surfaces of mirror-symmetric (010)-oriented slabs at $E=15\; {\rm meV}$ and both arcs are interrupted into three segments due to hybridization with a set of trivial surface bands. By studying the spin texture of surface states, we find the two Fermi arcs are strongly spin-polarized but in opposite directions, which can be detected by spin-polarized ARPES measurements. Our theoretical study of quasiparticle interference (QPI) for a nonmagnetic impurity at the Bi site also reveals several features related to the Fermi arcs. Specifically, we predict that the spin polarization of the Fermi arcs leads to a bifurcation-shaped feature only in the spin-dependent QPI spectrum, serving as a fingerprint of the Weyl nodes.
\end{abstract}
\date{\today}
\maketitle


\section{Introduction}
Despite being predicted more than 90 years ago as an elegant special solution of the relativistic wave equation, Weyl fermions have never been observed in nature.  However, just as the sun was setting on the Weyl fermion, they were observed as a quasiparticle in the electronic structure of TaAs, thereby pushing Weyl physics back to the vanguard and invigorating the search for other exotic relativistic particles within solids. Weyl fermions appear in condensed matter systems when a single Dirac cone is split into two by the breaking of either time-reversal \cite{Wan11,Xu11,Wang16,Morali19,Liu19,Belopolski19} or inversion symmetry \cite{Weng15, Xu15, Lv15, Liu16, Liu17,Huang16,Tamai16}. This process generates a pair of cone-like features with opposite chirality defined by their Berry curvature \cite{RMP18}. 
As a direct consequence, anomalous surface states appear connecting Weyl nodes of opposite topological charge, providing an experimentally accessible fingerprint of the underlying non-trivial band topology \cite{RevModPhys06,Inoue16,Zheng16,Batabyal16,Sessi17,Gyenis16,Chang16,Morali19}.

Recently, magnetic Weyl semi-metals have been gaining interest for providing a new platform to study the interplay between chirality, magnetism, correlation, and topological order, thus opening up new routes to novel quantum states, spin polarized chiral transport \cite{Hosur13, Juan17}, and exotic optical phenomena \cite{Morimoto16,Zyuzin17,Sirica19,Sirica20}. Compared with inversion breaking Weyl materials, magnetic Weyl materials have the advantage of reducing the minimum number of allowed Weyl nodes from four to two. This allows simpler band structures thereby facilitating clearer comparisons with theoretical predictions, and more robust applications in spintronics \cite{Zyuzin12,Vazifeh13,Li16} and quantum computation \cite{Meng12,Hosur14,He17,Yan20}. Another advantage is the magnetic field tunability which provides us a strong tool on controlling the band structure and related electromagnetic functionality.

To date very few magnetic Weyl materials have been synthesized, demonstrating an urgent and important need for more theoretical material predictions and robust growth protocols. Only very recently have ${\rm Co_3Sn_2S_2}$ and ${\rm Co_2MnGa}$ been reported as possible magnetic Weyl semimetals with six Weyl nodes and nodal lines, respectively \cite{Morali19,Liu19,Belopolski19}. This is far from enough to explore the wide range of possible exotic states. Interestingly, recent experimental reports of a pronounced negative magnetoresistance and observed band inversion in cerium monopnictides~\cite{Guo17,Kuroda18,PCanfield:2020} and the geometrically frustrated Shastry-Sutherland lattice GdB$_4$~\cite{Shon19} could provide clues to a possible new material family harboring Weyl physics.


In this paper, we argue  that CeBi in the ferromagnetic (FM) state is a magnetic Weyl semimetal using first-principles LDA+$U$ calculations and we predict robust quasiparticle interference (QPI) signatures of spin polarized Fermi arcs directly accessible by spin resolved scanning tunneling spectroscopy (STS). By aligning the Ce magnetic moments along the $c$-axis, we find two pairs of Weyl nodes on the zone diagonal along the $k_z$-axis at $15\;{\rm meV}$ above and $100\;{\rm meV}$ below the Fermi level. Then by examining a (010)-oriented slab which is mirror symmetric, two Fermi arcs of opposite spin polarization are observed on both the top and bottom of surfaces, stretching across the Brillouin zone. These Fermi arcs then hybridize with a set of trivial surface states producing a spin vortex encircling the $\bar{M}$ point in the Brillouin zone. Finally, by considering a weak potential scatterer at the ${\rm Bi}$ site on the surface, we predict several features directly related to the scattering among the Fermi arcs in the QPI spectra. Specifically, there is a bifurcation-shaped feature originating from the scattering between the two symmetric outer segments of the spin-up Fermi arc. We propose this feature as a direct indicator of the Weyl nodes for future experiments. Moreover, due to the isolation of the Weyl nodes near the Fermi level, we find CeBi to be a robust platform for theoretical and experimental analysis of Weyl physics with minimal interference from trivial bulk states.

\section{Method}
First-principles band structure calculations were carried out within density functional theory framework using the generalized-gradient approximation (GGA) as implemented in the all-electron code WIEN2K~\cite{wien2k19}, which is based on the augmented-plane-wave + local-orbitals(APW+lo) basis set. Spin-orbit coupling was included in the self-consistency cycles. Furthermore, the Hubbard Coulomb interaction on Ce-4$f$ electrons was incorporated to ensure a fully localized Ce-$4f$ state, which also gives a total magnetic moment of $2\mu_B$/Ce consistent with earlier studies oncerium monopnictides~\cite{Liechtenstein:1994,Guo17,Brinda19}. Although these compounds exhibit several field-induced magnetic phases, a ferromagnetic-like state can be stabilized using state-of-the-art cryogenic STM in 3D vector magnetic fields above 4 Tesla~\cite{Guo17,Brinda19,Trainer17}. The topological analysis was performed by employing a real-space tight-binding model Hamiltonian, which was obtained by using the wien2wannier interface~\cite{Jan10}. Bi-$6p$ and Ce-$5d$ states were included in generating Wannier functions. It is worth mentioning that, the DFT+$U$ method for a non-magnetic phase will still leave $f$-electron pinned around the Fermi level, which is completely different from the spin-polarized case~\cite{Tutchton20}. Therefore, the DFT+$U$ approach is not appropriate to describe strongly correlated materials in a paramagnetic state, for which one needs to resort to more powerful methods like DFT+DMFT.~\cite{Ryu20}

To identify the presence of a pair of Weyl nodes in a given band structure we must check if the band crossings are topologically trivial or non-trivial. That is, we must calculate the Berry curvature at and surrounding the nodal points to see if the field is divergent or not. To do so, we calculate the Berry curvature $\Omega_{n,\alpha\beta}(\mathbf{k})$ as implemented in the WannierTools package\cite{Wu18,wang2006ab,Xiao10},  for a given band $n$ and momenta $\mathbf{k}$ as 
\begin{align}
\Omega_{n,\alpha\beta}(\mathbf{k})=
-2  \text{Im} \sum_{m\neq n}
\frac{    v_{nm,\alpha}(\mathbf{k})    v_{mn,\beta}(\mathbf{k})   }{    \left[ \varepsilon_{m\mathbf{k}}-\varepsilon_{n\mathbf{k}} \right]^2 }
\end{align}
where $\varepsilon_{n\mathbf{k}}$ is the eigenvalue of Hamiltonian $H$ for band $n$ and momenta $\mathbf{k}$ and the matrix elements of the Cartesian velocity 
operators 
are given by 
\begin{align}
 v_{nm,\alpha}(\mathbf{k})
 =
\braket{  \phi_{n\mathbf{k}} \left|  \frac{\partial \hat{H}(\mathbf{k})}{\partial k_\alpha}  \right| \phi_{m\mathbf{k}} }.
\end{align}
Finally the Berry curvature vector is given by
\begin{align}
\Omega_{n,\gamma}(\mathbf{k}) = \varepsilon_{\alpha\beta\gamma} \Omega_{n,\alpha\beta}(\mathbf{k})
\end{align}
 where $\varepsilon_{\alpha\beta\gamma} $ is the Levi-Civita tensor. 

The surface electronic structure was calculated using both the direct diagonalization and the iterative Green’s function method~\cite{Sancho85,supp}. The QPI spectra was obtained within the $T$-matrix approximation as given by:
\begin{eqnarray}\label{EQ:QPI}
{\rm QPI}({\textbf q},\omega) &&= i(2\pi)^{-1}\!\!\! \int d^2{\textbf k}(2\pi)^{-2}[B({\textbf q},\omega)\!\!-\!\!B^*(-{\textbf q},\omega)], 
\end{eqnarray}
where $B({\textbf q},\omega) = \text{Tr}[G({\textbf k}+{\textbf q},\omega){\rm \bf T}(\omega)G({\textbf k},\omega)]$ with ${\textbf q}$ and $\omega$ the scattered wave vector  and quasiparticle energy, respectively. $G$ is the spin-dependent matrix surface Green's function composed of 6 Bi-$6p$ and 10 Ce-$5d$ orbitals, and ${\rm \bf T}$ is the scattering $T$-matrix containing all multiple scattering effects off a single-site impurity. Since the infinite sum of multiple scattering terms form a geometric series, the $T$-matrix can be compactly written as ${\rm \bf T}(\omega)=\left[ {\rm \bf I}-{\textbf V_{imp}}(2\pi)^{-2}\int d^2{\textbf k}G({\textbf k},\omega) \right]^{-1}{\textbf V_{imp}}$,
where ${\textbf V_{imp}}$ is the impurity potential. Here we considered a Born approximation of the impurity scattering such that ${\textbf T}={\textbf V_{imp}}$ and further chose 
${\textbf V_{imp}}$ to be a constant diagonal matrix with non-zero elements only for $p$ ($d$) orbitals for a Bi (Ce) impurity site. 
 Our general methodology for QPI prediction in CeBi is justified by the good agreement between calculated and experimental STS data on the (001)-oriented slab, shown in detail in the Supplemental Material (SM)~\cite{supp}.

\begin{figure}[t]
\begin{center}
\includegraphics[clip = true, width =\columnwidth]{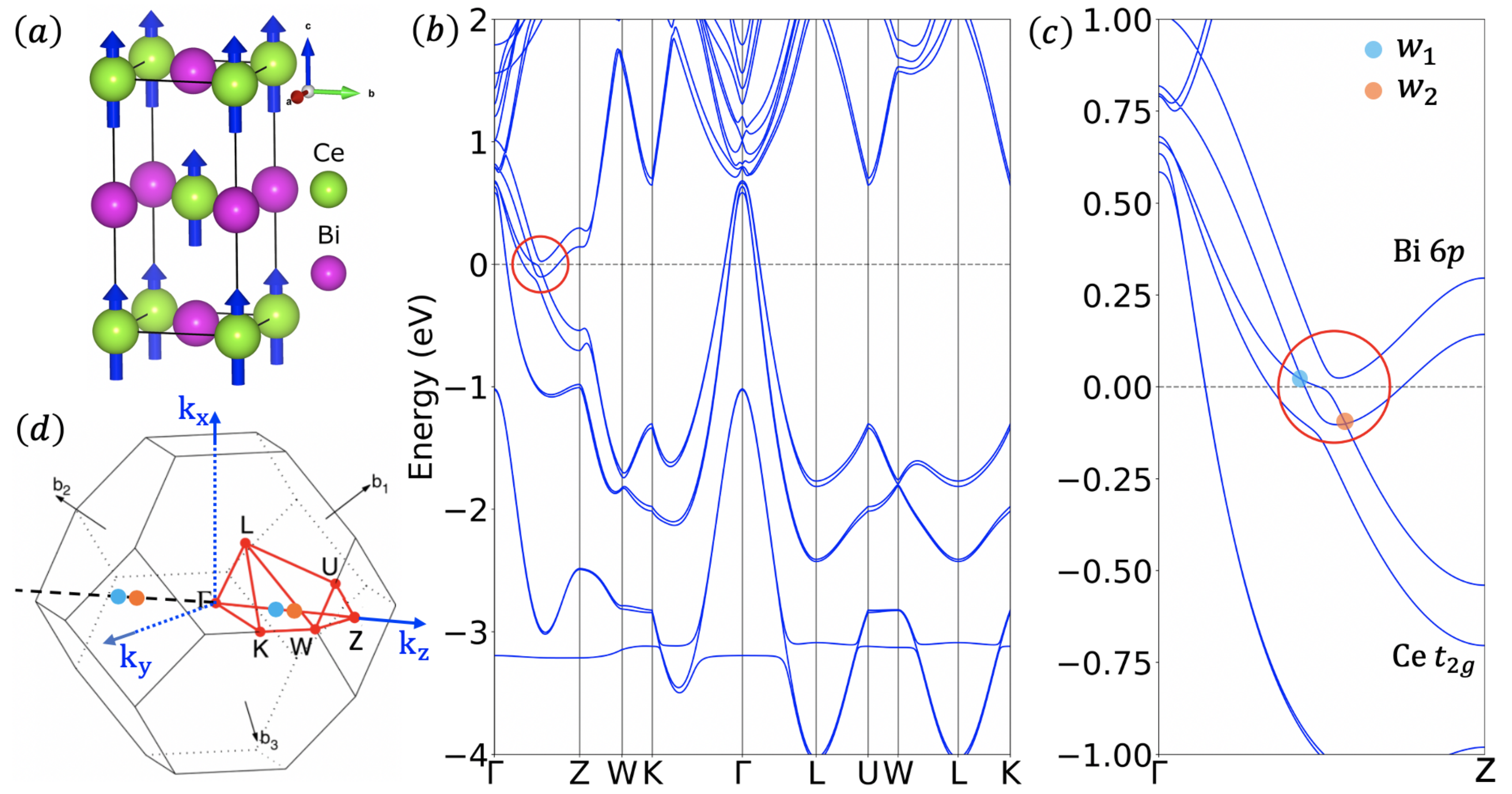}
\caption{(a) Crystal structure of CeBi with the spin polarization along the $c$-axis direction. Here we display a body-centered tetragonal cell for an otherwise FCC crystal structure of CeBi in bulk. (b) DFT electronic band structure of CeBi in the ferromagnetic state with a red circle highlighting a pair of Weyl nodes near the Fermi level. (c) Same as (b) except a zoom-in around Fermi level along $\Gamma-Z$ with the character of various bands indicated. (d) FCC Brillouin zone of FM CeBi with the various high-symmetry points marked. The locations of the Weyl nodes is given by teal (orange) dots denoting the Weyl node as a source (sink) of Berry curvature.}
\label{fig1}
\end{center}
\end{figure}

\section{band structure and Weyl nodes} 
Figure~\ref{fig1} (left panel) shows the crystal structure of CeBi in the FM phase along with its corresponding face-centered-cubic (FCC) Brillouin zone, where various high symmetry lines are marked. The middle panel presents the electronic band dispersions evaluated over the full set of high symmetry points. The energy bands near the Fermi level are mainly of Bi-$6p$ and Ce-$5d$ character. Moreover, the moment-carrying localized Ce-$4f$ states, which provides  an effective Zeeman exchange field, are seen as a flat  band at $3$ eV binding energies. We choose $U=7.9\; {\rm eV}$ and $J=0.69\;{\rm eV}$ in our {\it ab initio} calculations to make the Ce-$4f$ energy level consistent with the corresponding state in CeSb as reported by photoemission spectroscopy~\cite{Jang19}. Moreover, the non-trivial topological nature of low-energy states in CeBi are not sensitive to the value of $U$ once  Ce-4$f$ electrons are quite localized in the ferromagnetic phase.

The most important feature of the band structure is the pair of Weyl nodes that appears along $\Gamma - Z$ at the Fermi level. These two Weyl nodes are formed by the crossing of ${\rm Bi}$-$6p$ and ${\rm Ce}$-5$d$ $t_{2g}$ orbital character bands that have been spin split by the internal Zeeman exchange field [Fig.~\ref{fig1} (right panel)]. The Weyl nodes $w_1$ and $w_2$ are located at $(0,0,0.22)\frac{4\pi}{a}$ and $(0,0,0.28)\frac{4\pi}{a}$ in the Brillouin zone and at $15\;{\rm meV}$ above and $100\;{\rm meV}$ below the Fermi level, respectively. Here $a$ is the lattice constant of the cubic conventional cell and hereafter we will choose $4\pi/a=1$. Due to mirror symmetry in the $xy$-plane, a duplicate pair of Weyl nodes are found along the $-k_z$ axis, named as $\bar{w}_1$ and $\bar{w}_2$. The locations of these four Weyl nodes are marked in the Brillouin zone [Fig.~\ref{fig1} (left panel)] by  the teal (orange) dots, where the teal (orange) color indicates a Weyl node as a source (sink) of Berry curvature. Importantly, since the Weyl nodes in CeBi are isolated near the Fermi level, the experimental signatures and theoretical analysis will be less obstructed as compared to ${\rm Co_3Sn_2S_2}$ and ${\rm Co_2MnGa}$.

\begin{figure}[t]
\begin{center}
\includegraphics[clip = true, width =1.0\columnwidth]{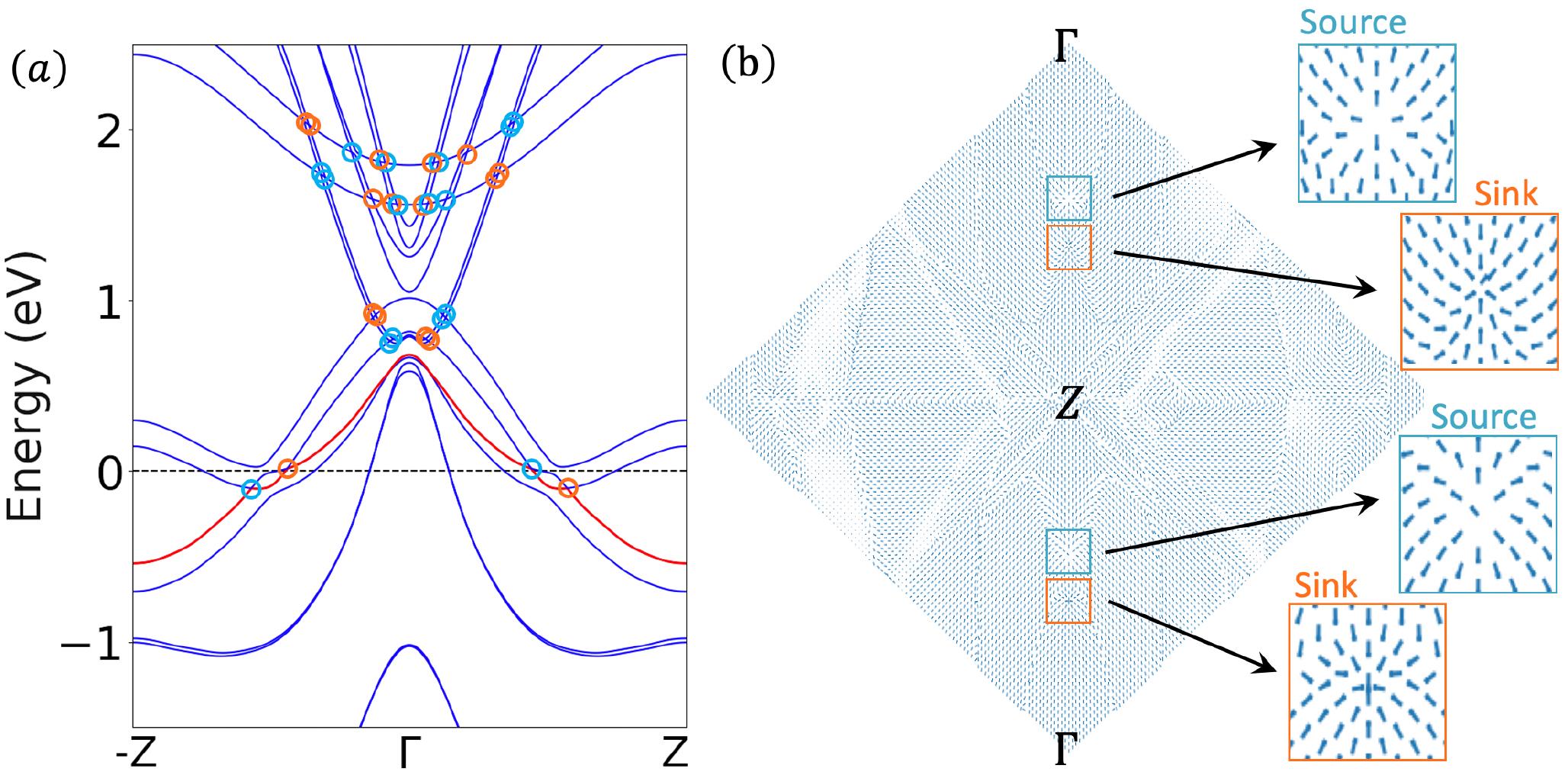}
\caption{Electronic band structure of ${\rm CeBi}$ in the FM phase along $-Z-\Gamma-Z$ in the Brillouin zone. The band that passes through the Weyl nodes near the Fermi level is shown in red. (b) Berry curvature vector field in the $xz$-plane for the red band in (a). The Weyl nodes are marked in teal (orange) indicating they are a source (sink) of Berry flux. }
\label{fig2}
\end{center}
\end{figure}

Figure~\ref{fig2}(a) shows $w_1$ and $w_2$ near the Fermi level along the $-Z - \Gamma - Z$ cut in the Brillouin zone. The band that passes through the Weyl nodes is marked in red. Figure~\ref{fig2}(b) shows the corresponding Berry curvature vector field in the $xz$-plane of the first Brillouin zone for the red band in Fig. \ref{fig2}(a). Two clear divergences in the vector field are seen along $\pm Z-\Gamma$ at the Weyl nodes, with one node as a source (teal) and the other a sink (orange) of topological charge. This confirms the topologically non-trivial nature of these Weyl band crossings. Furthermore, we identified 16 additional pairs of Weyl node existing above the Fermi energy, each marked in Fig. \ref{fig2}(a) by an orange or teal circle denoting its topological charge.

\section{surface spectral functions}

Due to the presence of Weyl nodes near the Fermi level we expect the emergence of a pair of Fermi arcs connecting nodes of opposite topological charge when CeBi is cleaved along a given crystal plane exposing a surface. However, the situation is more subtle in magnetic Weyl materials. Since the internal magnetic moments on the Ce atoms point in the $c$-axis direction, the irreducible group elements are different for a (010)-oriented surface compared to those for a (001)-oriented surface. Specifically, the (001)-oriented surface breaks the $xy$ mirror symmetry and $k_z$ is no longer a good quantum number, whereas a (010)- or (100)-oriented surface preserves the symmetry. 
Furthermore,  a (001)-oriented surface will not reveal Fermi arcs since the pairs of Weyl nodes with opposite chirality are projected to the same point on the surface, thus cancelling each other, while a (010) or (100)-oriented surface is expected to produce the Fermi arcs.  Here we focus on the (010)-oriented surface and examine the resulting electronic structure and effect of the Fermi arcs on the QPI spectrum.

\begin{figure}[t]
\begin{center}
\includegraphics[clip = true, width =1.0\columnwidth]{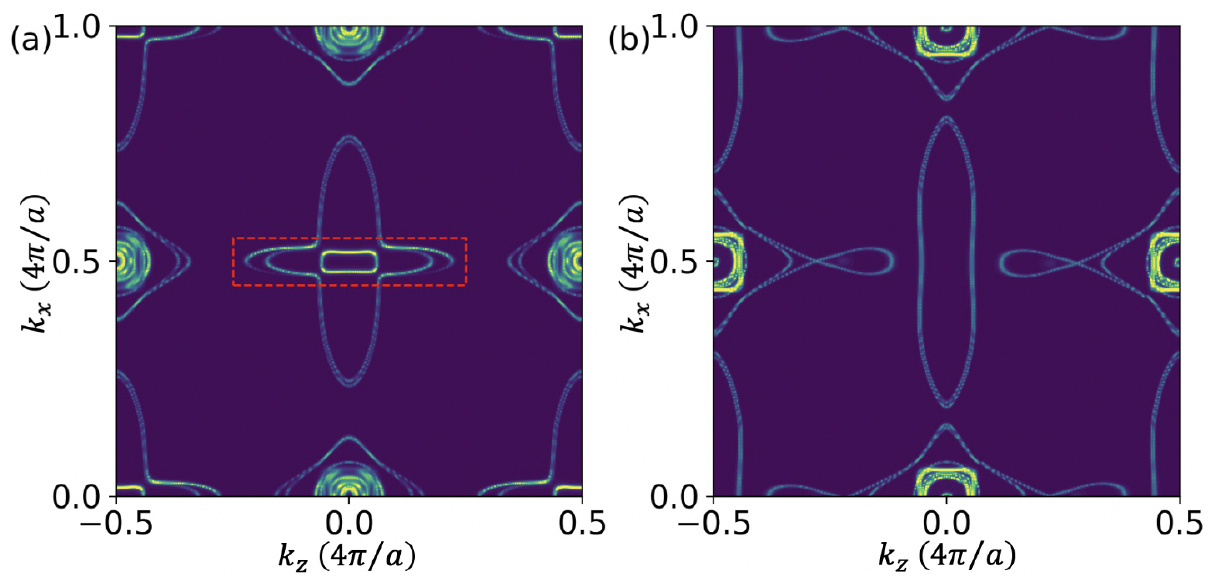}
\caption{Surface projected spectral function of the bottom layer of 10-layer $(010)$-oriented slab (a) and the corresponding bulk (b).}
\label{fig3}
\end{center}
\end{figure}

To distinguish the topologically non-trivial surface states and the trivial surface bands derived from the bulk, we consider the band structure of a 10-layer (010)-oriented slab, and also  the bulk with supercell containing 10 layers of CeBi along the same direction. The spectral functions for the bottom layer of both cases are presented in Fig. \ref{fig3}. By comparing the two panels, we can see that all bands in Fig. \ref{fig3}(a) except for the Fermi arcs [highlighted in the red frame] have the corresponding bulk bands in Fig. \ref{fig3}(b). This indicates that the Fermi arcs are the non-trivial topology induced edge states. The massless energy bands corresponding to Fermi arcs are shown in the supplementary \cite{supp}.

\begin{figure}[t]
\begin{center}
\includegraphics[clip = true, width =\columnwidth]{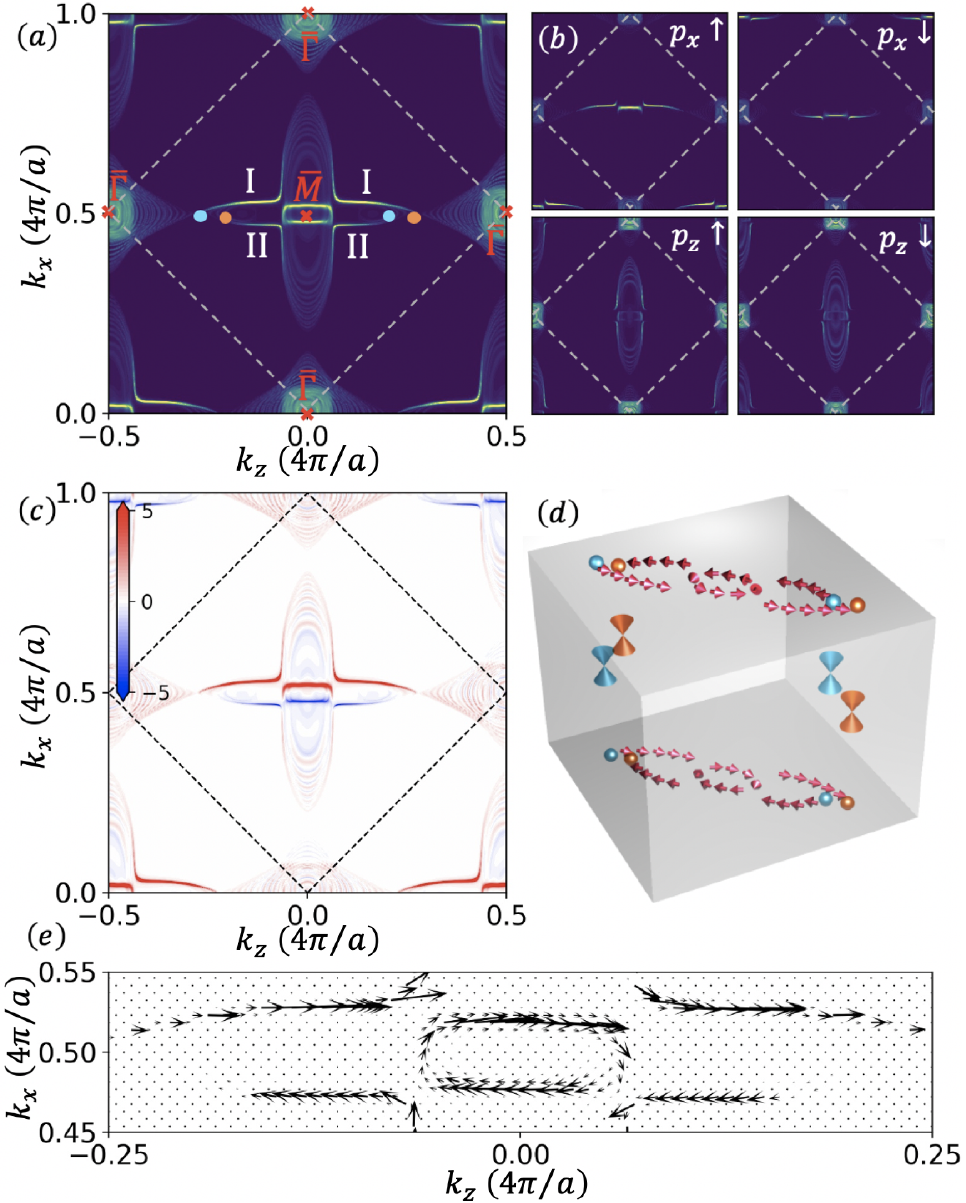}
\caption{((a) Surface spectral function of a (010)-oriented surface of FM CeBi. The dash lines draw the boundary of the surface first Brillouin zone. ${\rm I}$ and ${\rm II}$ denotes the two Fermi arcs connecting the Weyl nodes. (b) Spin- and orbital-resolved surface spectral functions. (c) Spin polarization $\langle S_z \rangle$ in the Brillouin zone. (d) Schematic of the spin texture of various Fermi arcs in the Brillouin zone. (e) Calculated spin texture of the Fermi arcs. }
\label{fig4}
\end{center}
\end{figure}

Figure~\ref{fig4}(a) shows the surface spectral function $A_{\text{surf}}(\mathbf{k},\omega)$ for an effective 64-layer (010)-oriented slab in the surface reciprocal space, centered at the $\bar{M}$ point, for a constant-energy cut  of $\omega$ corresponding to $w_1$. Two clear Fermi arcs are seen connecting projections of the Weyl nodes $w_1$ to  $\bar{w}_1$ and $w_2$ to  $\bar{w}_2$, labeled as arc I and arc II, respectively. By breaking down the spectral function into its orbital components, we find the Fermi arcs to be mainly composed of Bi $p_x$ and $p_y$, and Ce $t_{2g}$ orbitals. The dominant Bi $p_x$ and weaker Bi $p_z$ contribution, along with their spin dependence, is given in Fig. \ref{fig4}(b).

Interestingly, as the two Fermi arcs traverse across the Brillouin zone they are intersected by and hybridize with two majority $p_z$ character trivial bands extending along the $k_x$ direction [Fig.~\ref{fig4}(a)]. The avoided crossing between these two sets of bands yields a nearly continuous outer cross shaped Fermi surface and a rectangular pocket surrounding $\bar{M}$. Moreover, additional trivial surface states are observed around the $\bar{\Gamma}$ point, but they do not interact with the arcs. 

Figure \ref{fig4}(c) shows $\langle s_z({\rm \bf k},\omega) \rangle=-\frac{1}{\pi}{\rm ImTr}[\sigma_3G({\rm \bf k},\omega)]$ in the first Brillouin zone for a constant-energy cut  of $\omega=w_1$, where red (blue) indicates the strength of the positive (negative) $\sigma_3$ projection of the surface spectral function. Here we find arc I and arc II to be oppositely spin polarized, which is also reflected in Fig.~\ref{fig4}(b).  This is the consequence of the spin-momentum locking effect around the $\bar{M}$ point which originates from the strong spin-orbital coupling in CeBi. Moreover, the strength of polarization is generically greater for arc I compared to arc II, due to the FM order along the $c$-axis. 

Figure~\ref{fig4}(d) and (e) show the full spin texture of the Fermi arcs in the (010)-oriented slab. We note several important features: (i) Arc I is strongly polarized along $+z$, whereas arc II is strongly polarized along $-z$. (ii) Due to inversion symmetry, the spin polarization for a given $(k_z, k_x)$ on the bottom surface is equivalent to the spin polarization on the top surface at $(-k_z, -k_x)$. (iii) The Fermi surface pocket surrounding $\bar{M}$ forms a spin vortex in the momentum space, owing to strong spin-orbit coupling. (iv) The spin-polarization vector lies completely in the $(k_z,k_x)$ plane, where $\langle s_y \rangle$ is strictly zero. The absence of any out-of-plane spin polarization $S_y$ is guaranteed by the combined symmetry of time-reversal symmetry $T$ and twofold rotational symmetry along the $y$ axis $C_{2y}$, even though $T$ or $C_{2y}$ itself is broken. This symmetry is satisfied the Hamiltonian of the (010) slab with $C_{2y}TH(k_x, k_z)T^{-1}C_{2y}^{-1}=H(k_x,k_z)$. Since $T$ flips all spin components and reverses all crystal momenta, and $C_{2y}$ only flips $S_x,S_z$ and reverses $k_z, k_x$, the combined transformation $C_{2y}T$ maps $(k_z, k_x)$ to itself and reverses $S_y$. Because $(C_{2y}T)^2=1$, there is no Kramers degeneracy, implying $C_{2y}T|k_z,k_x\rangle$ and $|k_z,k_x\rangle$ can only differ by a phase, forbidding different $S_y$. Therefore, the expectation value of $S_y$ must be strictly zero throughout the Brillouin zone.

\begin{figure}[t]
\begin{center}
\includegraphics[clip = true, width =\columnwidth]{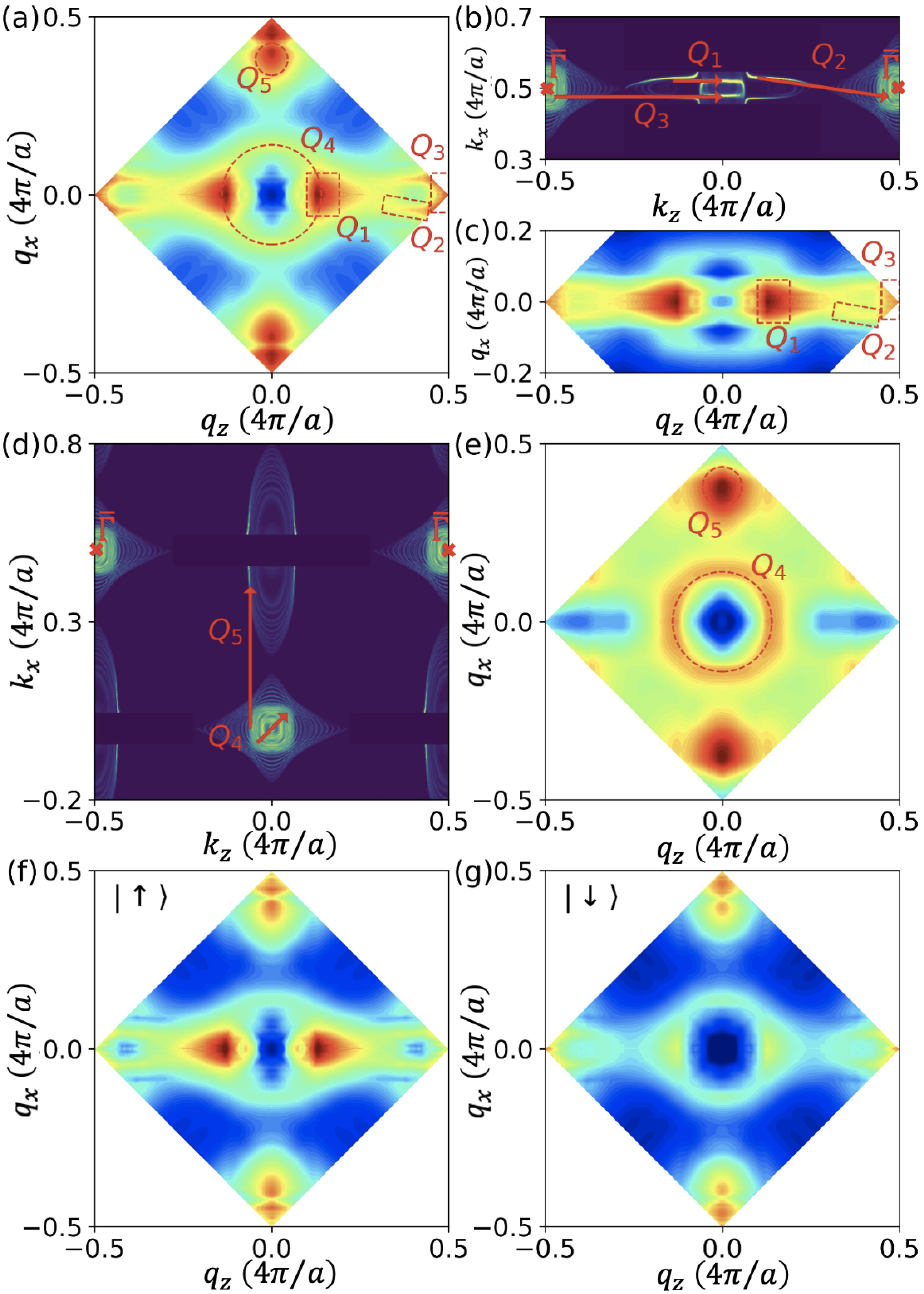}
\caption{(a) QPI spectra for a (010)-oriented surface of FM CeBi with the dominant scattering vectors indicated. (b) Surface spectral function with the trivial $p_z$ states removed with various characteristic scattering transitions marked. (c) QPI spectra corresponding to the surface spectral function in (b). (d) Surface spectral function with Fermi arcs and spin vortex removed, with dominant scattering vectors indicated. (e) QPI spectra corresponding to the surface spectral function in (d). QPI spectra of the spin-up (f) and spin-down channel (g) corresponding to the surface spectral function in Fig. \ref{fig4}(a).}
\label{fig5}
\end{center}
\end{figure}

\begin{figure}[t]
\begin{center}
\includegraphics[clip = true, width = \columnwidth]{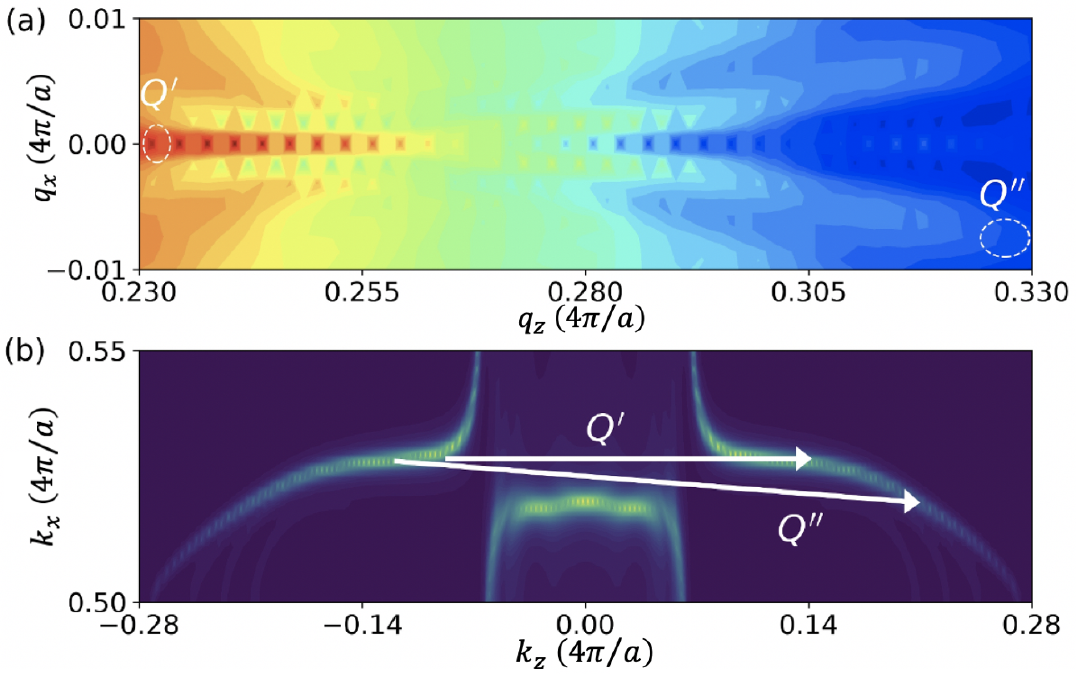}
\caption{(a) Zoom-in view of the QPI spectra in Fig. \ref{fig5}(a) highlighting the bifurcated feature corresponding to direct Fermi arc-Fermi arc scattering. (b) Close-up of the surface spectral function in Fig. \ref{fig4}(a) with scattering vectors $Q’$ and $Q"$ connecting the two outer segments of the spin-up Fermi arc.}
\label{fig6}
\end{center}
\end{figure}

\section{Quasiparticle Intereference}
{\it  Bi-site impurity} -- At first we present our theoretical study on the (010)-oriented surface QPI derived from scattering between these surface states due to an isotropic nonmagnetic impurity at Bi site. Figure~\ref{fig5}(a) shows the QPI spectra in the first Brillouin zone for a constant energy cut of $\omega=w_1$. Firstly, we notice that almost all $\mathbf{q}=0$ scattering is suppressed by destructive interference between the various multiorbital scattering channels. This phenomena is not present in the simple joint density of state description since the phase of wavefunctions is neglected. But because ${\rm QPI}(0,\omega)\propto \int d^2{\rm \bf k} {\rm ImTr}[G({\textbf k},\omega)T({\textbf k},\omega)G({\textbf k},\omega)]$, the off diagonal components of the matrix Green's function contribute to the QPI signal. Moreover, the sign of the imaginary part of the product is highly sensitive to the relative sign between real and imaginary parts of the Green function and ${\textbf k}$ in the Brillouin zone. Therefore, even after integrating over the full zone the results may be quite small. A similar cancellation takes place for scattering at finite ${\rm \bf q}$, which leads to a quite broad and smooth QPI spectrum despite sharp features in the spectral function.

In Fig.~\ref{fig5}(a) the five main features in the QPI pattern, corresponding to strong scattering between the surface states, are labeled with $Q_1$ through $Q_5$ and are highlighted by red dash circles and rectangles. In order to pinpoint the origin of the these strong features, we isolate various sections of the surface spectral function by masking out the Green function in other sections and amplify their effect on the QPI spectra. If we only consider the scattering between the Fermi arcs, spin vortex, and bands surrounding $\bar{\Gamma}$ [Fig.~\ref{fig5}(b)] we find the $Q_1,\ Q_2$ and $Q_3$ features persist in the QPI [Fig.~\ref{fig5}(c)]. Therefore, we find the momentum transfer $Q_1$  to connect the outer segment of the Fermi arc to the pocket surrounding $\bar{M}$. $Q_2$ scatters electrons from the outer segments of Fermi arcs to the trivial bands around $\bar{\Gamma}$. Because the Fermi arc is curved, a curved bridge-like dispersion is seen in the $Q_2$ rectangle.  $Q_3$ transfers momentum between the Fermi pocket to the trivial bands around $\bar{\Gamma}$. If we now consider the spectral function in the region away from the Fermi arcs and spin vortex [Fig. \ref{fig5}(d)], we find the remaining Fermi sheets clearly produce the remaining features $Q_4$ and $Q_5$ [Fig. \ref{fig5}(e)]. $Q_4$ comes from the internal scattering of the bands around $\bar{\Gamma}$, whereas the $Q_5$ scattering vector connects $p_z$ dominant states to those surrounding $\bar{\Gamma}$.

Since the surface spectral function exhibits a strong spin texture, it is reasonable to expect a highly spin dependent QPI. Figures \ref{fig5}(f)-(g) show the up and down-spin projection of the QPI spectra. These two patterns are quite similar except for scattering momenta related to $Q_1$. Since $Q_1$ connects Fermi arcs of the same $+z$ spin polarization, the intensity in the spin-up QPI channel is stronger than that in the spin-down channel. This stark difference in the spin-up and spin-down QPI maps provides a direct experimentally accessible prediction of the underlying spin-polarized Fermi arcs.

\begin{figure}[t]
\begin{center}
\includegraphics[clip = true, width =  \columnwidth]{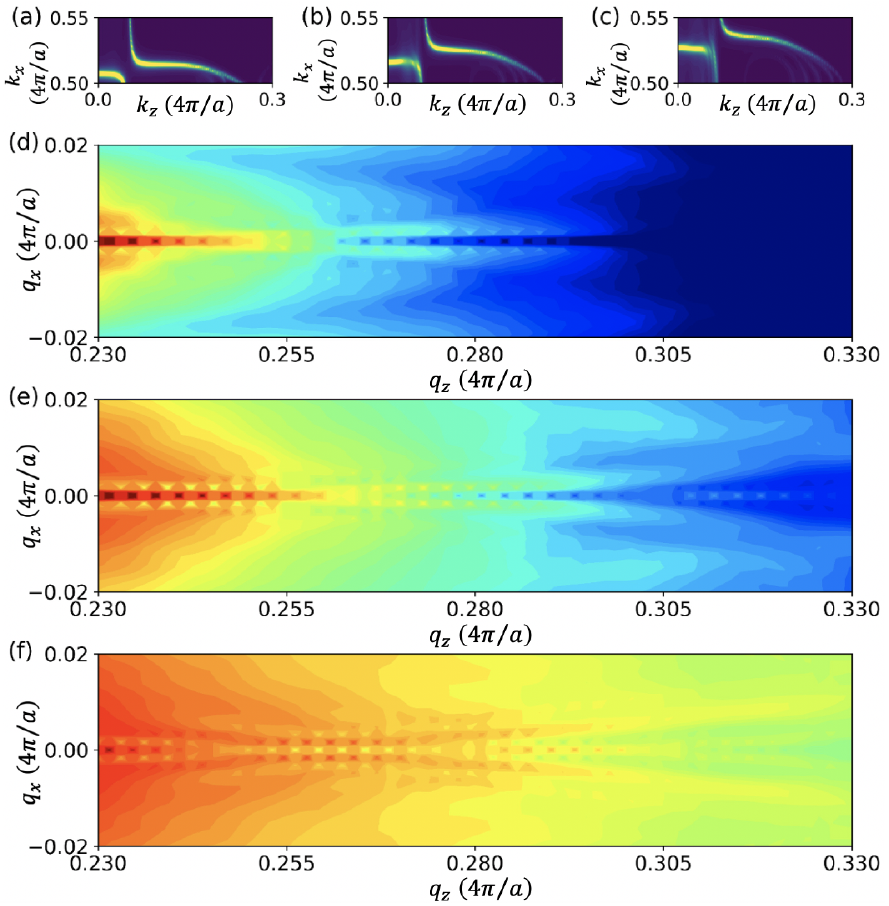}
\caption{Spectral function around one Fermi arc at (a) $-50$  meV, (b) 0 meV and (c)  50 meV. QPI spectra in the same regime as Fig.~\ref{fig4}(a) at (d) $-50$ meV, (e) 0 meV and (f) 50 meV.}
\label{fig7}
\end{center}
\end{figure}

\begin{figure*}[t]
\begin{center}
\includegraphics[clip = true, width = 1.8 \columnwidth]{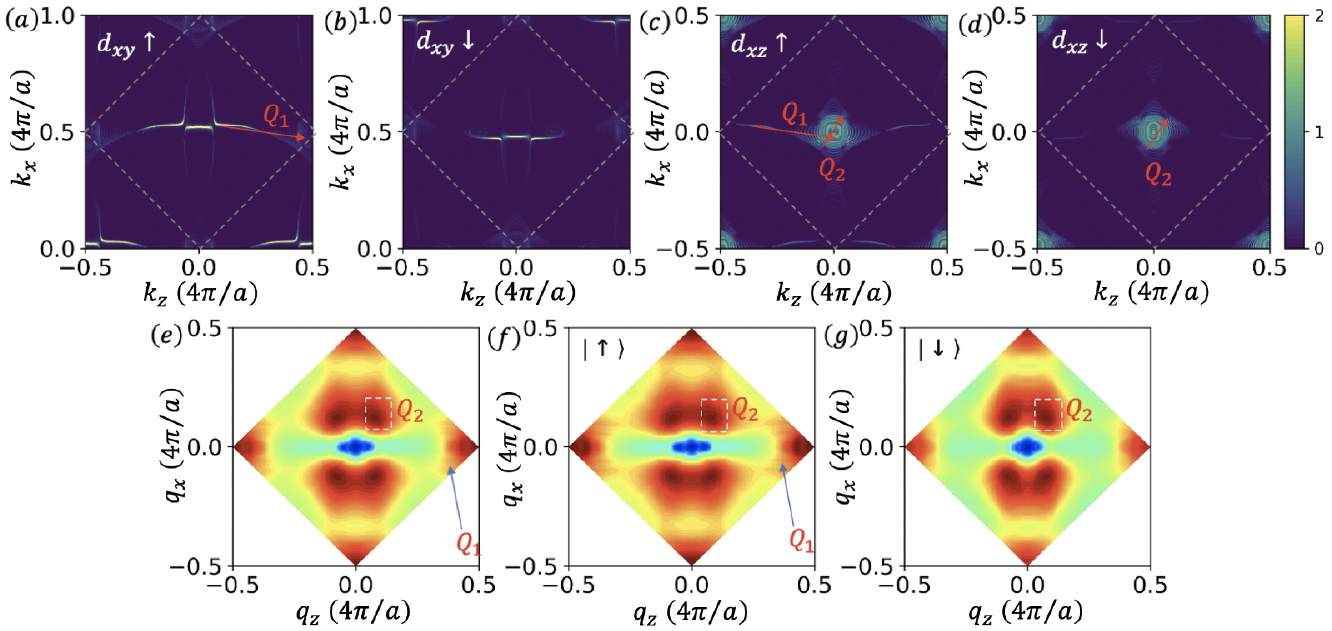}
\caption{Spectral functions of (a) spin-up $d_{xy}$, (b) spin-down $d_{xy}$, (c) spin-up $d_{xz}$, and (d) spin-down $d_{xz}$. (e) total, (f) spin-up and (g) spin-down QPI spectrum.}
\label{fig8}
\end{center}
\end{figure*}

Thus far, we have covered all the strong scattering transitions in the QPI map. However, the scattering between the two outer Fermi arcs appears to be missing. There should be features in the QPI spectra corresponding to this scattering channel. Indeed, upon close inspection, a red flat tail can be seen at $(q_z,q_x)=(\pm0.25,0)$ in Figs. \ref{fig5}(a), \ref{fig5}(c) and \ref{fig5}(f) corresponding to this `missing' scattering pathway. 
Figure~\ref{fig6}(a) shows a zoomed in section of the QPI map given in Fig.~\ref{fig5}(a) displaying a dispersing bifurcating feature. The trunk, which is the tail mentioned above, extends from $q_z=0.23$ to about $q_z=0.26$, and bifurcates into two branches. The intensity of the two branches decreases with increasing $|q_z|$, but the branch between $q_z=0.26$ to $q_z=0.33$ is still clearly seen. This bifurcation shaped feature originates from electrons scattered between the shoulder-like portion  (denoted by $Q^\prime$) and the slope (marked by $Q^{\prime\prime}$) of the Fermi arc, as shown in Fig. \ref{fig6}(b). The enhanced scattering along $q_x=0$ originates from pure shoulder-shoulder $Q^\prime$ scattering spanning between the mirrored sections of the Fermi arcs. For $q_z>0.26$, $Q^{\prime\prime}$ momentum transfers becomes dominant, tracing out the curvature of the Fermi arc. The intensity of the QPI decreases increased $Q^{\prime\prime}$ due to the reduction in the nesting between the shoulder and the curving Fermi arc, therefore, generating the two bifurcating branches in the QPI spectrum.

Figure~\ref{fig7} shows the energy dependence of the Fermi arc I and the bifurcation feature. From panel (a)-(c), we see that as energy increases from $-50 {\rm meV}$ to  $50 {\rm meV}$, the Fermi arc becomes more curved due to the energetic distance from the Weyl node\cite{supp}. This manifests in QPI as a growing angle of the bifurcation branches as shown in Fig. \ref{fig7}(d)-(f).

{\it Ce-site impurity} --  Now we replace the ${\rm Bi}$ impurity with a ${\rm Ce}$ impurity and mark the difference in the surface spectral function and QPI. Here, the $d$-orbitals play the key role in the ${\rm Ce}$ impurity induced scattering. We find two noteworthy scattering vectors labeled $Q_1$ and $Q_2$ shown in Fig. \ref{fig8}. $Q_1$ connects the Fermi arcs with trivial bands around $\Gamma$. Additionally, because the Ce magnetic moments are along the positive $z$ direction, the spectral function for spin-up $d$ orbitals are more intense than the spin-down $d$ orbitals. As a result, the $Q_1$ spike-like feature in the total and spin-up QPI spectrum is shown in Fig. \ref{fig8}(e) and \ref{fig8}(f), whereas this feature is not clear in Fig. \ref{fig8}(g). The momentum transfer $Q_2$, which is similar to $Q_4$ in the case of ${\rm Bi}$ impurity, comes from the internal scattering of the bands around the $\Gamma$ point. This scattering channel is strong for both spin-up and spin-down channels, which makes it a strong feature in all QPI maps.

The features displayed for a ${\rm Ce}$ impurity are not as rich as those found for a ${\rm Bi}$-site scatter because the existence of a dense set of Ce-$5d$ bands surrounding the $\bar{\Gamma}$ point at the Fermi level that generate a significant trivial band scattering  background, thus, washing out most features related to Fermi arc scattering in the QPI.

\vspace{1cm}

\begin{figure*}[t]
\begin{center}
\includegraphics[clip = true, width = 1.4 \columnwidth]{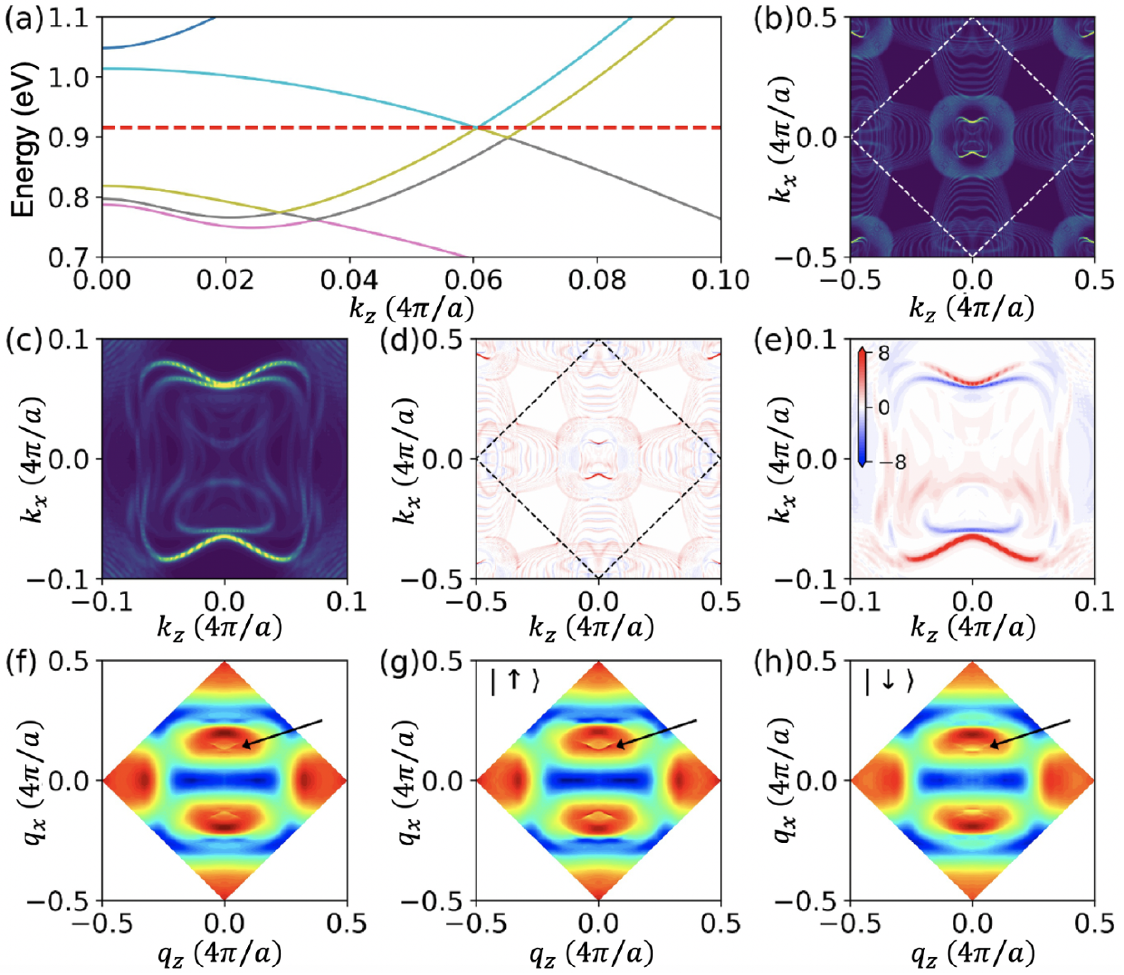}
\caption{(a) Theoretical electronic band structure around the Weyl nodes at $0.915\;\text{eV}$. Surface projected spectral function for $\omega=0.915\;\text{eV}$ in (b) the first Brillouin zone and (c) zoom-in around $\Gamma$ showing the Fermi arcs. $\langle S_z \rangle$  in (d) the first Brillouin zone and (e) zoom-in around $\Gamma$. (f) total, (g) spin-up and (h) spin-down QPI spectra. }
\label{fig9}
\end{center}
\end{figure*}

\section{QPI for Higher Energy Weyl Node}
In the previous sections, we have studied the surface states and QPI spectra originating from near the Fermi level of a $(010)$-oriented slab. Here, we study the surface projected spectral function and QPI map of the $(010)$-oriented slab for a pair of Weyl nodes above the Fermi level at $0.915\;{\rm eV}$, as indicated by the red dashed line in Fig. \ref{fig9}(a). We chose this specific pair of Weyl nodes since they are quite isolated from other intruding bands, which tend to complicate the observation of the Fermi arcs. This Weyl node and the one just below it are both sources of Berry flux as indicated in Fig. \ref{fig2}(a). The surface spectral function at this energy is shown by Fig. \ref{fig9}(b), where two clear wing-like features are found near $\Gamma$ along the $k_x$ axis. Zooming-in on these features [Fig. \ref{fig9}(c)], it is easy to see that they are Fermi arcs which connect the Weyl nodes. Figure \ref{fig9}(d) shows the $\sigma_3$ projection of the spectral function $\langle S_z \rangle$ in the first Brillouin zone, where the red (blue) intensity indicates the strength of the spin-up (spin-down) polarization. Figure \ref{fig9}(e) shows a close-up showing a clear spin texture of the Fermi arc. Similar to our earlier discussion, the scattering between the Fermi arcs with different spin textures can lead to interesting features in the total and spin resolved QPI spectra, as shown in Fig. \ref{fig9}(f), \ref{fig9}(g) and \ref{fig9}(h). Because the spin-up Fermi arcs are curved and the spin-down arcs are relatively flat, the QPI map yields curved and flat features in the spin-up and spin-down QPI spectra, respectively, as indicated by the black arrows. The total QPI also displays a similar curved feature as shown in Fig. \ref{fig9}(f).

\section{Conclusion}
In summary, we have found theoretically that ${\rm CeBi}$ is a magnetic Weyl semimetal with two pairs of Weyl nodes close to the Fermi level. The induced two Fermi arcs have opposite spin polarization directions on each surface of a mirror symmetric slab. A nonmagnetic impurity at ${\rm Bi}$ site can lead to quasiparticle interference pattern with features arising from the Fermi arcs. There is a novel bifurcation shaped feature available in spin-up but absent in spin-down QPI pattern, which is a unique property of this material and can be verified with spin-resolved scanning tunneling probes. Our calculations provide a baseline, from which further experimental and theoretical studies may be initiated to examine the other aspects of Weyl physics in CeBi and other magnetic Weyl materials.

\begin{acknowledgments}
{\it Acknowledgments} -- We thank  Jenny Hoffman for useful discussions. This work was carried out under the auspices of the U.S. Department of Energy (DOE) National Nuclear Security Administration under Contract No. 89233218CNA000001. It was supported by the Center for the Advancement of Topological Semimetals, a DOE BES EFRC. C.L. was supported by LANL LDRD Program. Additional support was provided in part by the Center for Integrated Nanotechnologies, a DOE BES user facility, in partnership with the LANL Institutional Computing Program for computational resources. C.E.M. was supported by the Swiss National Science Foundation under fellowship No. P2EZP2\_175155 and P400P2\_183890.
\end{acknowledgments}

%

\end{document}


\title{Supplemental Material for "Prediction of Spin Polarized Fermi Arcs in Quasiparticle Interference of CeBi"}

\author{Zhao Huang}
\affiliation{Theoretical Division, Los Alamos National Laboratory, Los Alamos, New Mexico 87545, USA}

\author{Christopher Lane}
\affiliation{Theoretical Division, Los Alamos National Laboratory, Los Alamos, New Mexico 87545, USA}
\affiliation{Center for Integrated Nanotechnology, Los Alamos National Laboratory, Los Alamos, New Mexico 87545, USA}

\author {Chao Cao}
\affiliation{Department of Physics, Hangzhou Normal University, Hangzhou 310036, China}

\author {Guo-Xiang Zhi}
\affiliation{Department of Physics, Zhejiang University, Hangzhou 310013, China}

\author{Yu Liu}
\affiliation{Department of Physics, Harvard University, Cambridge, Massachusetts 02138, USA}

\author{Christian E. Matt}
\affiliation{Department of Physics, Harvard University, Cambridge, Massachusetts  02138, USA}


\author{Brinda Kuthanazhi}
\affiliation{Ames Laboratory, Iowa State University, Ames, Iowa 50011, USA}
\affiliation{Department of Physics and Astronomy, Iowa State University, Ames, Iowa 50011, USA}

\author{Paul C. Canfield}
\affiliation{Ames Laboratory, Iowa State University, Ames, Iowa 50011, USA}
\affiliation{Department of Physics and Astronomy, Iowa State University, Ames, Iowa 50011, USA}

\author{Dmitry Yarotski}
\affiliation{Center for Integrated Nanotechnology, Los Alamos National Laboratory, Los Alamos, New Mexico 87545, USA}

\author{A. J. Taylor}
\affiliation{Center for Integrated Nanotechnology, Los Alamos National Laboratory, Los Alamos, New Mexico 87545, USA}

\author{Jian-Xin Zhu}
\email{jxzhu@lanl.gov}
\affiliation{Theoretical Division, Los Alamos National Laboratory, Los Alamos, New Mexico 87545, USA}
\affiliation{Center for Integrated Nanotechnology, Los Alamos National Laboratory, Los Alamos, New Mexico 87545, USA}

\date{\today}

\maketitle


\section{Formula of Quasiparticle Interference}

In this section we derive the expression for the quasiparticle interference for the surface layer of a slab with two atoms A and B in a primitive cell. The location of all atoms on the surface is given by an in-plane vector ${\rm \bf r_i}$, where $i$ {\it even} denotes the A sublattice and $i$ {\it odd} gives the B sublattice. The Green's function of the entire slab with an impurity at site ${\rm \bf r_1}$ (type B) is given by
\begin{eqnarray}
\Sigma_{j'}[i\omega_n\delta_{ij'}{\rm\textbf I}-{\rm\textbf h_{ij'}}]{\rm\textbf G(j',j)=\delta_{ij}{\rm\textbf I}}+{\rm\textbf V_{imp}}\delta_{i1}{\rm\textbf G(1,j)},
\end{eqnarray}
where $i, j, j'$ index the atom locations throughout the entire slab, ${\rm\textbf h_{ij'}}$ is the hopping matrix connecting orbitals in unit cell $i$  to those in unit cell $j'$, and ${\rm \textbf V_{imp}}$ is the impurity potential matrix. Moreover, ${\rm \textbf V_{imp}}$ is block diagonal in atomic species, with only a 6 by 6 nonzero block for B-type vacancy only composed of 3 Bi-$6p$ orbitals plus spin. This means if the A site is composed of 5 Ce-5$d$ orbitals plus spin degrees of freedom, the overall dimension of ${\rm \bf I,\ h, G}$ and ${\rm \textbf V_{imp}}$ can grow rapidly for increased number of layers. Then by defining the Green's function in the pristine case
\begin{eqnarray}
\Sigma_{j'}[i\omega_n\delta_{ij'}{\rm\textbf I}-{\rm\textbf h_{ij'}}]{\rm\textbf G_0(j',j)=\delta_{ij}{\rm\textbf I}},
\end{eqnarray}
we can write the full greens function in terms of the ${\rm \bf T_{imp}}$ matrix,
\begin{eqnarray}
{\rm\textbf G(i,j)}={\rm\textbf G_0(i,j)}+{\rm\textbf G_0(i,1)}{\rm\textbf T_{imp}}{\rm\textbf G_0(1,j)},
\end{eqnarray}
where  ${\rm \bf T_{imp}}$ is obtained as
\begin{eqnarray}
{\rm\textbf T_{imp}}={\rm\textbf V_{imp}}
(I-{\rm\textbf G_0(1,1)}{\rm\textbf V_{imp}})^{-1}.
\end{eqnarray}
If we only consider the first Born approximation, where $V_{imp} \ll 1$, the deviations from ${\rm\textbf G_0}$ can be written as
\begin{eqnarray}
\delta{\rm\textbf G(i,i)} ={\rm\textbf G_0(i,1)} {\rm\textbf V_{imp}}{\rm\textbf G_0(1,i)} \;.
\end{eqnarray}
This determines the corresponding local density of states  for sites A and B,
\begin{eqnarray} 
&& \rho(2i)=-\frac{1}{\pi}{\rm ImTr}[{\rm\textbf G_0(2i,1)} {\rm\textbf V_{imp}}{\rm\textbf G_0(1,2i)}], \label{rho2i}\\
&& \rho(2i+1)=-\frac{1}{\pi}{\rm ImTr}[{\rm\textbf G_0(2i+1,1)} {\rm\textbf V_{imp}}{\rm\textbf G_0(1,2i+1)}]. \label{rho2i1}
\end{eqnarray}

Since we only wish to examine the QPI spectrum relevant to STM measurements, we restrict $\textbf G_0$ to just surface atomic sites. Therefore, we define the surface local density of states for sites A and B as
\begin{eqnarray} 
&& \rho_s(2i)=-\frac{1}{\pi}{\rm ImTr}[{\rm\textbf G_{s0}(2i,1)} {\rm\textbf V_{p}}{\rm\textbf G_{s0}(1,2i)}], \label{rhos2i}\\
&& \rho_s(2i+1)=-\frac{1}{\pi}{\rm ImTr}[{\rm\textbf G_{s0}(2i+1,1)} {\rm\textbf V_{p}}{\rm\textbf G_{s0}(1,2i+1)}], \label{rhos2i1}
\end{eqnarray}
where $G_{s0}$ is the surface Green function with a dimension of  $10\times 6$ or $6\times 10$ $(6\times 6)$ for site A (B), and ${\rm\textbf V_{p}}$ is the nonzero $6\times 6$ block of ${\rm\textbf V_{imp}}$. Since we have periodic boundary conditions in the $xy$ plane of the slab geometry, we are able to perform Fourier transform to obtain the quasiparticle interference spectral function at the surface as 
\begin{eqnarray}
&& \rho_{sA}({\rm\textbf q})=\sum_i \rho_s(2i) \exp(-i{\rm\textbf q}\cdot {\rm\textbf r_{2i}}), \label{Aq1}\\
&& {\rm\textbf G_{s0}(2i,1)}=\langle d_{2i}p_1^\dagger \rangle=\sum_{\rm\textbf k}{\rm\textbf G_{s0}^{dp}({\rm\textbf k})}\exp(i{\rm\textbf k}\cdot ({\rm\textbf r_{2i}}-{\rm\textbf r_1}))\ \;\text{with}\; \ {\rm\textbf G_{s0}^{dp}}({\rm\textbf k})= \langle d_{\rm\textbf k}p_{\rm\textbf k}^\dagger \rangle,  \label{Aq2} \\
&& {\rm\textbf G_{s0}(1,2i)}=\langle p_1d_{2i}^\dagger \rangle=\sum_{\rm\textbf k}{\rm\textbf G_{s0}^{pd}({\rm\textbf k})}\exp(-i{\rm\textbf k}\cdot ({\rm\textbf r_{2i}}-{\rm\textbf r_1}))\ \; \text{with}\; \ {\rm\textbf G_{s0}^{pd}}({\rm\textbf k})= \langle p_{\rm\textbf k}d_{\rm\textbf k}^\dagger \rangle,  \label{Aq3} \\
&& \rho_{sB}({\rm\textbf q})=\sum_i \rho_s(2i+1) \exp(-i{\rm\textbf q}\cdot {\rm\textbf r_{2i+1}}), \label{Bq1} \\
&& {\rm\textbf G_{s0}(2i+1,1)}=\langle p_{2i+1}p_1^\dagger \rangle=\sum_{\rm\textbf k}{\rm\textbf G_{s0}^{pp}({\rm\textbf k})}\exp[i{\rm\textbf k}\cdot ({\rm\textbf r_{2i+1}}-{\rm\textbf r_{1}})]\ 
\; \text{with} \; \ {\rm\textbf G_{s0}^{pp}}({\rm\textbf k})= \langle p_{\rm\textbf k}p_{\rm\textbf k}^\dagger \rangle, \label{Bq2}\\ 
&& {\rm\textbf G_{s0}(1,2i+1)}=\sum_{\rm\textbf k}{\rm\textbf G_{s0}^{pp}({\rm\textbf k})}\exp[-i{\rm\textbf k}\cdot ({\rm\textbf r_{2i+1}}-{\rm\textbf r_{1}})]. \label{Bq3}
\end{eqnarray}
Combining Eqs.~(\ref{rhos2i}), (\ref{Aq1}), (\ref{Aq2}) and (\ref{Aq3}), we arrive at
\begin{eqnarray}
\rho_{sA}({\rm\textbf q})&&=\sum_i -\frac{1}{\pi}{\rm ImTr}[{\rm\textbf G_{s0}(2i,1)} {\rm\textbf V_{p}}{\rm\textbf G_{s0}(1,2i)}] \exp(-i{\rm\textbf q}\cdot {\rm\textbf r_{2i}}) \\
&&=\frac{i}{2\pi}\sum_i \sum_{k,k'} {\rm Tr}[{\rm\textbf G_{s0}^{dp}({\rm\textbf k})} {\rm\textbf V_{imp}}{\rm\textbf G_{s0}^{pd}({\rm\textbf k}')}\exp[i({\rm\textbf k}-{\rm\textbf k}')\cdot ({\rm\textbf r_{2i}}-{\rm\textbf r_{1}})]-c.c.] \exp(-i{\rm\textbf q}\cdot {\rm\textbf r_{2i}}) \\
&&=\frac{i}{2\pi}\sum_{k} {\rm Tr}[{\rm\textbf G_{s0}^{dp}({\rm\textbf k})} {\rm\textbf V_{imp}}{\rm\textbf G_{s0}^{pd}({\rm\textbf k}-{\rm\textbf q})}-[{\rm\textbf G_{s0}^{dp}({\rm\textbf k})} {\rm\textbf V_p}{\rm\textbf G_{s0}^{pd}({\rm\textbf k}+{\rm\textbf q})}]^*] \exp(-i{\rm\textbf q}\cdot {\rm\textbf r_1}),
\end{eqnarray}
and combining  Eq. (\ref{rhos2i1}), (\ref{Bq1}), (\ref{Bq2}) and (\ref{Bq3}) we find
\begin{eqnarray}
\rho_{sB}({\rm\textbf q})&&=\sum_i -\frac{1}{\pi}{\rm ImTr}[{\rm\textbf G_{s0}(2i+1,1)} {\rm\textbf V_p}{\rm\textbf G_{s0}(1,2i+1)}] \exp(-i{\rm\textbf q}\cdot {\rm\textbf r_{2i+1}}) \\
&&=\frac{i}{2\pi }\sum_i \sum_{k,k'} {\rm Tr}[{\rm\textbf G_{s0}^{pp}({\rm\textbf k})} {\rm\textbf V_{p}}{\rm\textbf G_{s0}^{pp}({\rm\textbf k}')} \exp[i({\rm\textbf k}-{\rm\textbf k}')\cdot ({\rm\textbf r_{2i+1}}-{\rm\textbf r_{1}})]-c.c.] \exp(-i{\rm\textbf q}\cdot {\rm\textbf r_{2i+1}}) \\
&&=\frac{i}{2\pi}\sum_{k} {\rm Tr}[{\rm\textbf G_{s0}^{pp}({\rm\textbf k})} {\rm\textbf V_{p}}{\rm\textbf G_{s0}^{pp}({\rm\textbf k}-{\rm\textbf q})}-[{\rm\textbf G_{s0}^{pp}({\rm\textbf k})} {\rm\textbf V_{p}}{\rm\textbf G_{s0}^{pp}({\rm\textbf k}+{\rm\textbf q})}]^*] \exp(-i{\rm\textbf q}\cdot {\rm\textbf r_1}).
\end{eqnarray}
To make the above expression more computationally tractable, we drop the common factor $\exp(-i{\rm\textbf q}\cdot {\rm\textbf r_1})$ for $\rho_{sA}$ and $\rho_{sB}$. Additionally, since ${\rm\textbf V_{imp}}=V{\rm\textbf I}$ for an isotropic nonmagnetic impurity we find
\begin{eqnarray}
\rho_{sA}({\rm\textbf q})=\frac{i}{2\pi}V\sum_{k} {\rm Tr}[{\rm\textbf G_{s0}^{dp}({\rm\textbf k}+{\rm\textbf q})}{\rm\textbf G_{s0}^{pd}({\rm\textbf k})}]-{\rm Tr}[{\rm\textbf G_{s0}^{dp}({\rm\textbf k})} {\rm\textbf G_0^{pd}({\rm\textbf k}+{\rm\textbf q})}]^* \\
=\frac{i}{2\pi}V\sum_{k} {\rm Tr}[{\rm\textbf G_{s0}^{dp}({\rm\textbf k}+{\rm\textbf q})}{\rm\textbf G_{s0}^{pd}({\rm\textbf k})}]-{\rm Tr}[{\rm\textbf G_0^{dp}({\rm\textbf k})}{\rm\textbf G_{s0}^{pd}({\rm\textbf k}+{\rm\textbf q})}]^*
\end{eqnarray}
and
\begin{eqnarray}
\rho_{sB}({\rm\textbf q})
&&=\frac{i}{2\pi}V\sum_{k} {\rm Tr}[{\rm\textbf G_{s0}^{pp}({\rm\textbf k}+{\rm\textbf q})}{\rm\textbf G_{s0}^{pp}({\rm\textbf k})}]-{\rm Tr}[{\rm\textbf G_{s0}^{pp}({\rm\textbf k})} {\rm\textbf G_{s0}^{pp}({\rm\textbf k}+{\rm\textbf q})}]^* \\
&&=-\frac{1}{\pi}V\sum_{k} {\rm Tr Im}[{\rm\textbf G_{s0}^{pp}({\rm\textbf k}+{\rm\textbf q})}{\rm\textbf G_{s0}^{pp}({\rm\textbf k})}].
\end{eqnarray}
Finally, the total spectral function is 
\begin{eqnarray}
{\rm B}({\rm\textbf q})=\rho_{sA}({\rm\textbf q})+\rho_{sB}({\rm\textbf q}) \label{totalQPI}.
\end{eqnarray}
In many cases, the off-diagonal contribution $\rho_{sA}$ is much smaller than $\rho_{sB}$. It is a good approximation to just keep only $\rho_{sB}({\rm\textbf q}))$ in $B(\mathbf{q})$~\cite{chang16}. In our case, because $\rho_{sA}$ is comparable with $\rho_{sB}$ for some ${\rm \bf q}$,  we still take the Eq.~(\ref{totalQPI}) to calculate the QPI spectra.

\section{Surface Electronic Structure and QPI of $(001)$-Oriented Surface}
In the following we compare our calculated surface spectral function and corresponding theoretically predicted QPI intensity with experimental results on the (001)-oriented surface, which is orthogonal to the (010)-oriented surface presented in the main manuscript. Due to technical limitations -- the available in-plane magnetic field is limited to 3 Tesla -- our QPI calculations and predictions of Fermi arcs on the (010)-oriented surface cannot be directly confirmed by experiments. However, the good agreement between our calculations and experiments on the (001)-oriented surface confirms and validates our methodology and calculations. 

\begin{figure}[h!]
\begin{center}
\includegraphics[clip = true, width =0.75\columnwidth]{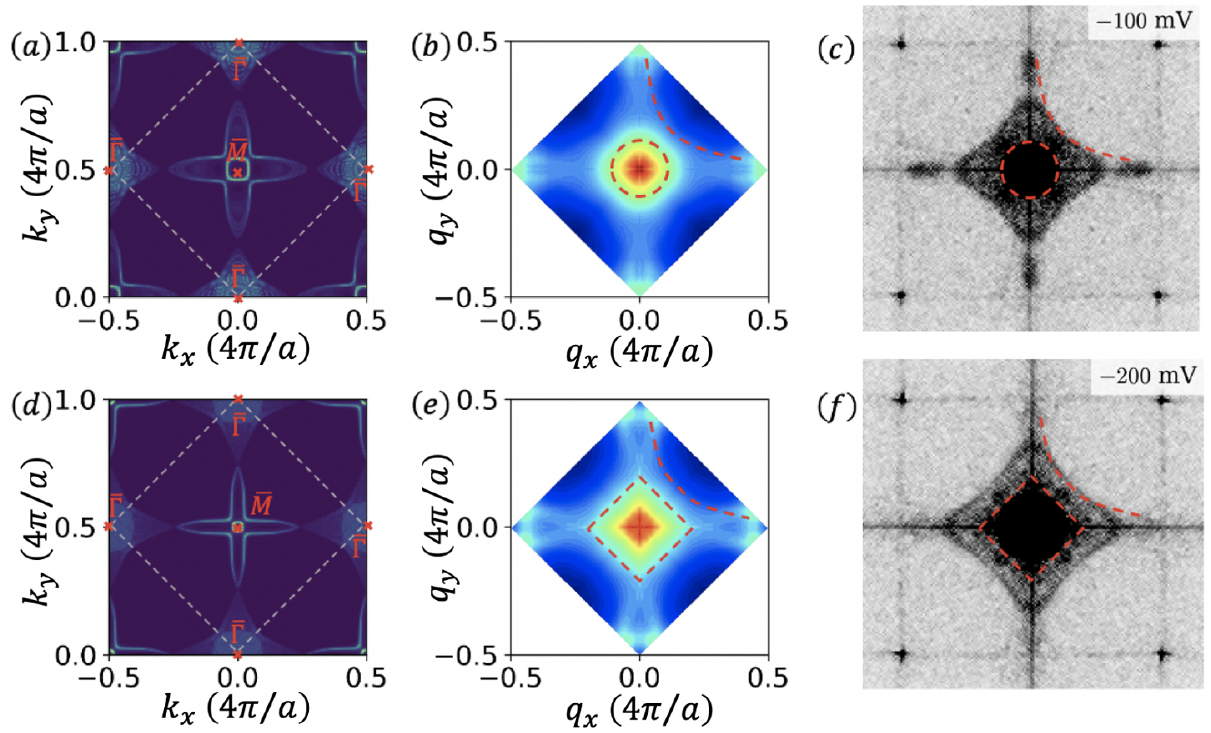}
\caption{Surface projected spectral function for (a) $\omega=-100\; {\rm meV}$ and (d) $\omega=-200\; {\rm meV}$. QPI spectra for a single Bi impurity for (b) $\omega=-100\; {\rm meV}$ and (e) $\omega=-200\; {\rm meV}$. Experimental QPI map measured at (c) $-100 {\rm mV}$ and (f) $-200 {\rm mV}$ bias voltages. Setup parameters for the QPI maps (c) and (f) are $V_s = 200$  mV, $R_J = 0.4\; \text{G}\Omega$, where $V_s$ denotes the sample bias and $R_J$ the tunneling junction resistance. The bias excitation amplitude was set to $V_\mathrm{rms} = 7.1$ mV.}
\label{exp}
\end{center}
\end{figure}

We have calculated the surface spectral function and the ${\rm Bi}$ vacancy induced QPI spectra at $\omega=-100\; {\rm meV}$ and $-200\; {\rm meV}$. Figure~\ref{exp} compares our theoretically obtained QPI spectra to experimental results. In our theoretical spectral functions [Fig. \ref{exp}(a) and \ref{exp}(d)] no Fermi arc is present due to the absence of $xy$-mirror symmetry in the $(001)$-oriented slab. Rather, a trivial cross-like Fermi surface pocket surrounding $\bar{M}$ and weak weight near $\bar{\Gamma}$ is present. In the corresponding QPI maps [Fig. \ref{exp}(b) and \ref{exp}(e)] the main scattering peak around $\mathbf{q}=0$ is circular for $\omega=-100\; {\rm meV}$, but transforms in to a rectangle for $\omega=-200\;{\rm meV}$. 

In our experiments, using scanning tunneling spectroscopy (STS) we measured QPI on the $(001)$-oriented surface of CeBi in the FM-like  state at 4 K (see Figs.~\ref{exp} (c), (f)).  
Single crystals of CeBi were grown out of a Bi-rich binary melt with an initial composition of Ce$_{26}$Bi$_{74}$~\cite{Brinda19}. Elemental Ce and Bi were placed in a 3-cap-Ta crucible~\cite{PCanfield:2020} and slowly cooled from 1200 \textdegree{}C to 940 \textdegree{}C over 60 hours.  Once at 940 \textdegree{}C the excess solution was decanted from the CeBi single crystals with the aid of a centrifuge~\cite{PCanfield:2020}.
CeBi samples were then cleaved in cryogenic, ultrahigh vacuum at 30 K before inserting them into the STM head. The Ce magnetic moments are fully aligned by a 9 Tesla external magnetic fields applied along the $c$-axis. By overlaying the key features of our theoretical results on top of the experimental QPI spectra [Fig. \ref{exp}(c) and \ref{exp}(f)] we find reasonable agreement. For the case of $\omega=-100\;{\rm meV}$, the theoretical and experimental results both have a circular high intensity peak around $\bar{\Gamma}$. Theory and experimental QPI is also consistent for the intense scattering near the corners of the Brillouin zone marked by a dashed red curve. For $\omega=-200\;{\rm meV}$, we find the same level of agreement between theory and experiment, where the size of the predicted and measured rectangle around $\bar{\Gamma}$ almost exactly the same. 

By considering an isotropic nonmagnetic impurity potential at the ${\rm Bi}$ site, we obtain theoretical results in good accord with the experimentally observed QPI. This suggests that Bi vacancies at the surface of CeBi behave as isotropic nonmagnetic impurities. Therefore, the T-matrix approximation $\rm \bf T=V_{imp}$ is reasonable and produces realistic results. 

\begin{figure}[h]
\begin{center}
\includegraphics[clip = true, width =0.6\columnwidth]{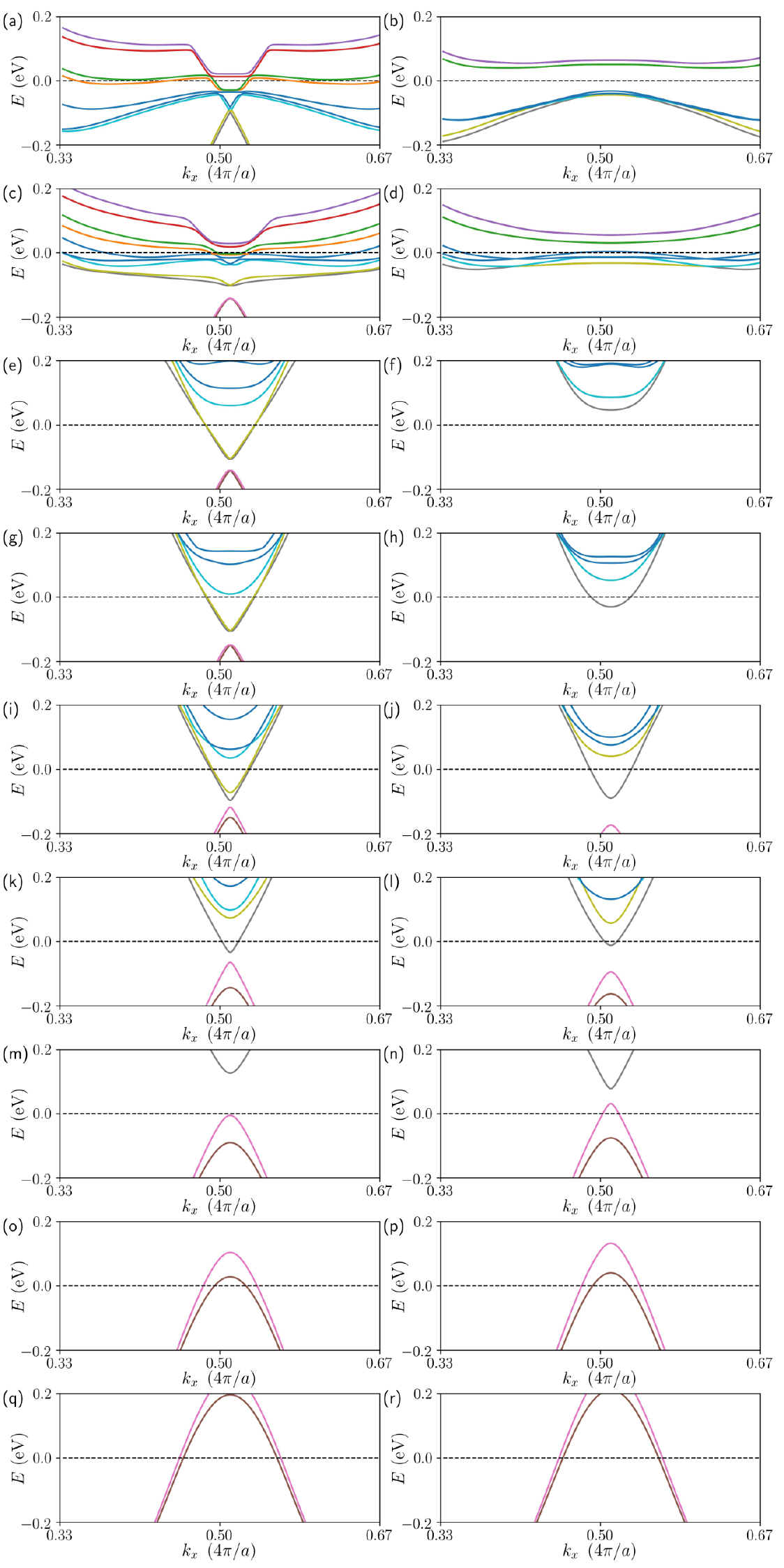}
\caption{Band structures of a 30-layer (010) slab with open and closed boundary condition in the $y$-direction for various values of $k_z$. $k_z=0,\ 0.05,\ 0.1,\ 0.15,\ 0.2,\ 0.25,\ 0.3,\ 0.35$ and $0.4$ for panels from the first to the last row. The boundary condition is open and closed for panels in the left and right column. }
\label{band}
\end{center}
\end{figure}

\begin{figure}[h!]
\begin{center}
\includegraphics[clip = true, width = 0.8 \columnwidth]{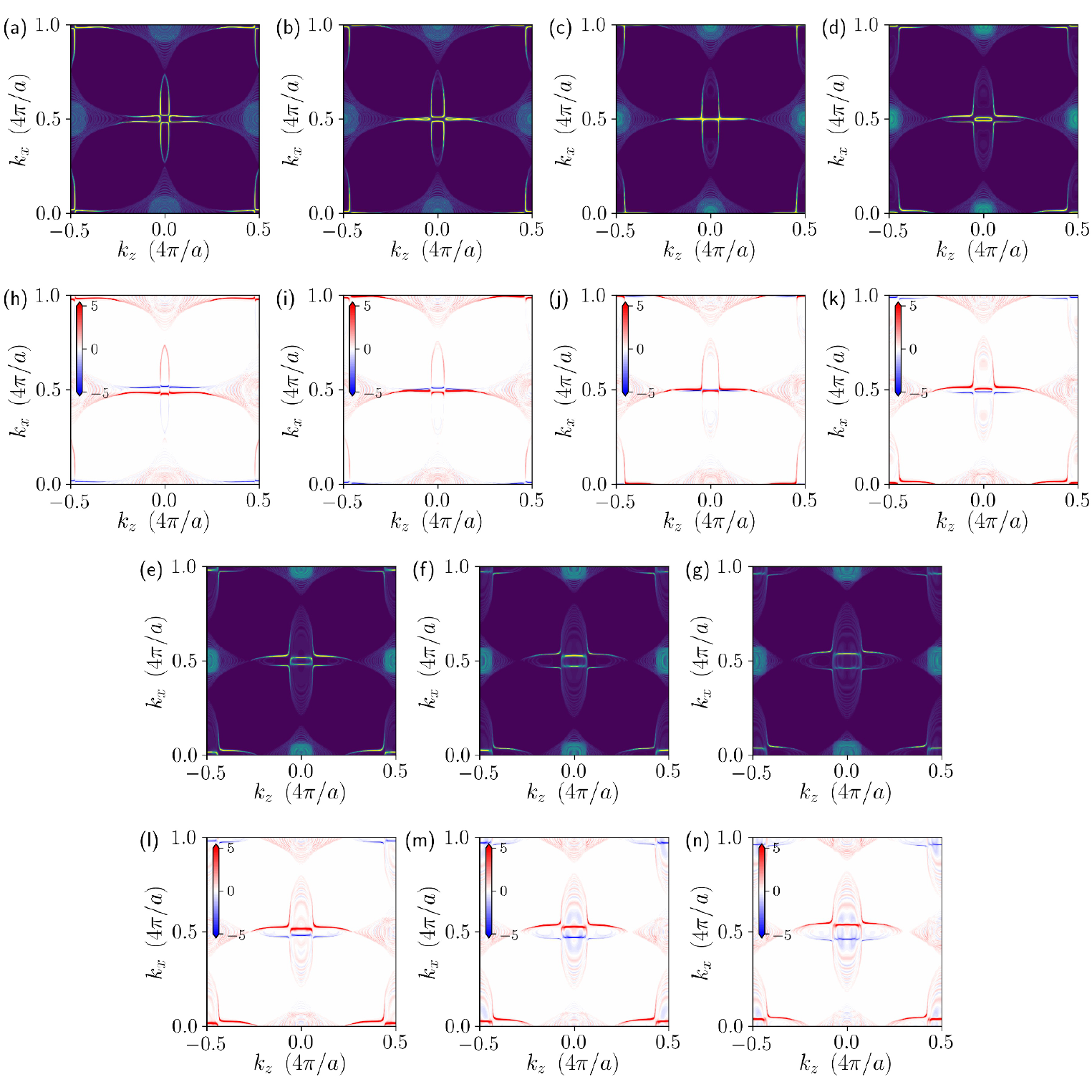}
\caption{(a) - (g) Surface spectral function of a (010)-oriented slab at $E=-0.2,\ -0.15,\ -0.1,\ -0.05,\ 0,\ 0.05$ and 0.1 eV. (h) - (n) Spectral functions for  spin polarization $\langle S_z \rangle$ at $E=-0.2,\ -0.15,\ -0.1,\ -0.05,\ 0,\ 0.05$ and 0.1 eV.  }
\label{spec}
\end{center}
\end{figure}

\vspace{-1cm}
\section{Topological nontrivial and trivial surface bands}

To show the band dispersions related to the Fermi arcs, we calculated the band structures of a 30-layer (010) slab with respect to $k_x$ for a list of $k_z$ as shown in Fig. \ref{band}. Both the open and closed boundary conditions are considered to distinguish the nontrivial surface band. We can see several features by comparing these panels. 1) Energy bands with linear dispersions are available in panel (a), (c), (e) and (g) but absent in (b), (d), (f) and (h). This means the existence of nontrivial massless bands at the open surface, which form the Fermi arcs. The bands have small gaps thus not perfectly massless because the finite depth of the slab. 2) We can see two linear-dispersion bands in (i) $(k_z=0.2)$ but only one in (k) $(k_z=0.25)$, which indicates a Weyl node between $k_z=0.2$ and $0.25$ because the 2D momentum plane between a pair of Weyl nodes has a nonzero Chern number \cite{RMP18}. 3) For the same reason, one linear-dispersion band in (k) but none in (m) indicates another Weyl node between $k_z=0.25$ and $0.3$.

The energy cuts on bands in Fig. \ref{band} correspond to the peaks in the map of spectral functions as shown in Fig. \ref{spec}. Specifically, the cuts on the bands with linear dispersion form the Fermi acrs. Fig. \ref{spec}(h)-(n) show the exchange of red and blue band (Fermi arcs) around the central area. This exchange is due to crossing of the massless linear bands shown in Fig. \ref{band}.

\bibliography{suppref}